# RADIATIVE PROTON CAPTURE ON $^{15}$N AT ASTROPHYSICAL ENERGY RANGE

S. B. Dubovichenko[1,2,*], N. Burtebaev[2,†], A. V. Dzhazairov-Kakhramanov[1,2,‡], D. K. Alimov[2]

[1]*V. G. Fessenkov Astrophysical Institute "NCSRT" NSA RK, 050020, Observatory 23, Kamenskoe plato, Almaty, Kazakhstan*
[2]*Institute of Nuclear Physics CAE MINT RK, 050032, str. Ibragimova 1, Almaty, Kazakhstan*
[*]*dubovichenko@mail.ru*
[†]*burteb@inp.kz*
[‡]*albert-j@yandex.ru*

The possibility of description of the experimental data for the astrophysical *S*-factor of the radiative proton capture on $^{15}$N at the energies from 50 to 1500 keV was considered in the framework of the modified potential cluster model with the classification of the orbital states according to Young tableaux. It was shown that, on the basis of the *M*1 and the *E*1 transitions from different states of the p$^{15}$N scattering to the ground state of $^{16}$O in the p$^{15}$N channel, it is quite succeed to explain general behavior of the *S*-factor in the considered energy range in the presence of two resonances.

## 1. Introduction

Continuing the study of the radiative proton capture reactions on light atomic nuclei, which are the part of different thermonuclear processes [1], let us stop on the capture reaction p$^{15}$N → $^{16}$Oγ at astrophysical energies. This process is a part of basic chain of thermonuclear reactions of the CNO cycle [2], which determine the formation of the Sun and stars at early stages of their evolution [2]

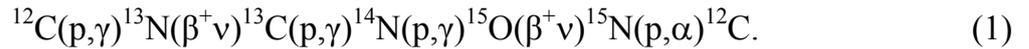

$$^{12}C(p,\gamma)^{13}N(\beta^+\nu)^{13}C(p,\gamma)^{14}N(p,\gamma)^{15}O(\beta^+\nu)^{15}N(p,\alpha)^{12}C. \qquad (1)$$

This chain has alternative reaction channel, which begins from the last reaction of the proton interaction with $^{15}$N [3]

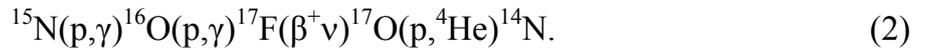

$$^{15}N(p,\gamma)^{16}O(p,\gamma)^{17}F(\beta^+\nu)^{17}O(p,^4He)^{14}N. \qquad (2)$$

Earlier in our works [4,5,6,7,8,9] the astrophysical *S*-factors and the total cross sections were considered for more than twenty radiative capture reactions of protons, neutrons and other charged particles on light atomic nuclei. The calculation methods, based on modified potential cluster model (PCM) of light nuclei with forbidden states (FSs) [10,11,12], are used for the analysis of such processes. The existence of the FS becomes clear on the basis of the classification of the orbital states of clusters according to Young tableaux [13]. In this approach the potentials of intercluster interactions for scattering processes [14,15,16,17,18,19,20,21] are constructed on the basis of the reproduction of elastic scattering phase shifts, taking into account their resonance behavior or spectrum of the final nucleus. The intercluster potentials are constructed on the basis of description of the binding energy and certain basic characteristics of the bound states (BSs) or the ground states (GSs) of nuclei in the cluster channels [14-22], for example, mean square radius and asymptotic constant (AC). The existence of the FSs and described criteria of construction of the potentials allow one to consider this PCM as modified (MPCM).

Choosing exactly the MPCM with FSs for considering of similar cluster systems and thermonuclear processes, is caused by the fact that in many light atomic nuclei the possibility of the formation of nucleon associations, i.e., clusters, and the degree of their isolation from each other are comparatively high. This is confirmed by the numerous experimental data and by the different theoretical calculations, obtained by different authors during the last fifty-sixty years [4-13].

## 2. The model and calculation methods

At first note that the classification of the orbital states of $^{15}$N according to Young tableaux was qualitatively considered in works [4-9,23]. Therefore, we have $\{1\} \times \{4443\} \rightarrow \{5443\} + \{4444\}$ [13] for the p$^{15}$N system in the frame of 1p-shell. The first of the obtained tableaux compatible with the orbital moments $L = 1$ and is forbidden as it contains five cells in the first row [13], and the second tableau is allowed and compatible with the orbital moment $L = 0$ [13]. Thereby, if to limit by the consideration of only lowest partial waves with orbital moment, it could be said that there is bound forbidden state in the $P$ wave potential, and the $S$ wave has no forbidden state. The allowed bound state in the $P$ wave corresponds to the GS of $^{16}$O and is at binding energy of the p$^{15}$N system of -12.1276 MeV [24]. Because the moment of $^{15}$N is equal to $J^{\pi}T = 1/2^-1/2$ [25] and for $^{16}$O we have $J^{\pi}T = 0^+0$, then its GS in the p$^{15}$N channel can be the $^3P_0$ state (representation in $^{(2S+1)}L_J$).

We regard the results on the classification of $^{16}$O by orbital symmetry in the p$^{15}$N channel as the qualitative one as there are no complete tables of Young tableaux productions for the systems with a number of nucleons more than eight [26], which have been used by us in earlier similar calculations [4-9,27,28]. At the same time, just on the basis of such classification, we succeeded with description of available experimental data on the radiative capture of protons and neutrons on $^{13}$C [29,30,31]. That is why here we will use similar classification of cluster states, which gives the certain number of forbidden and allowed states in different partial intercluster potentials. The number of such states determines the number of nodes of the wave function of cluster relative motion with the certain orbital moment $L$ [4-9].

The total radiative capture cross sections $\sigma(EJ)$ in the case of potential cluster model have the following form (see, for example, works [32] or [22,33]):

$$\sigma_c(NJ, J_f) = \frac{8\pi Ke^2}{\hbar^2 q^3} \frac{\mu}{(2S_1+1)(2S_2+1)} \frac{J+1}{J[(2J+1)!!]^2}$$
$$\times A_J^2(NJ,K) \sum_{L_i,J_i} P_J^2(NJ,J_f,J_i) I_J^2(J_f,J_i) \qquad (3)$$

where the next expression [32,33] is known for the orbital electric *EJ(L)* transitions:

$$P_J^2(EJ, J_f, J_i) = \delta_{S_i S_f} \left[(2J+1)(2L_i+1)(2J_i+1)(2J_f+1)\right] (L_i 0 J 0 | L_f 0)^2 \begin{Bmatrix} L_i & S & J_i \\ J_f & J & L_f \end{Bmatrix}^2 ,$$

$$A_J(EJ, K) = K^J \mu^J \left(\frac{Z_1}{m_1^J} + (-1)^J \frac{Z_2}{m_2^J}\right), \quad I_J(J_f, J_i) = \langle \chi_f | R^J | \chi_i \rangle . \quad (4)$$

Here $S_i$, $S_f$, $L_f$, $L_i$, $J_f$, $J_i$ – total spins, angular and total moments in initial (*i*) and final (*f*) channels; μ, *q* – reduced mass and wave number of the particles in initial channel; $m_1$, $m_2$, $Z_1$, $Z_2$ – masses and charges of the particles in initial channel; *K, J* – wave number and momentum of γ-quantum in final channel; $I_J$ –integral over wave functions of initial $\chi_i$ and final $\chi_f$ states, as functions of cluster relative motion with intercluster distance *R*; *N* – it is *E* or *M* transitions of *J* multipole order from the initial $J_i$ to the final $J_f$ nucleus state [33].

For consideration of the *M*1(*S*) magnetic transition, caused by the spin part of magnetic operator, it is possible to obtain expression [4-9] using expressions [34]:

$$P_1^2(M1, J_f, J_i) = \delta_{S_i S_f} \delta_{L_i L_f} \left[S(S+1)(2S+1)(2J_i+1)(2J_f+1)\right] \begin{Bmatrix} S & L & J_i \\ J_f & 1 & S \end{Bmatrix}^2 ,$$

$$A_1(M1, K) = \frac{e\hbar}{m_0 c} K \sqrt{3} \left[\mu_1 \frac{m_2}{m} - \mu_2 \frac{m_1}{m}\right], \quad I_J(J_f, J_i) = \langle \chi_f | R^{J-1} | \chi_i \rangle , \quad J = 1. \quad (5)$$

where, *m* is the mass of nucleus, $\mu_1$, $\mu_2$ are the magnetic moments of clusters the values of which are taken from [35] and [36].

Our computer program based on the finite-difference method (FDM) [37,38] was rewritten for carrying out of the calculation. The program was converted into the language Fortran-90 [4-9], which allows appreciably rise the computational speed and accuracy, and, for example, obtains more accurate values of the binding energy of nucleus in two-body channel [4-9]. The accurate value of the neutron mass [36] with the mass of $^{15}$N equals 15.000108 amu [39] were used in this calculations and constant $\hbar^2/m_0$ takes equal to 41.4686 MeV fm$^2$. Although, currently, this value is slightly out-of-date, we continue to use it for easing comparison of the last [4-9] and all earlier obtained results [22,33].

## 3. Interaction potentials and structure of resonance states

In the first place we have considered *E*1 transitions from the resonance $^3S_1$ scattering wave without FS at the energies up to 1.5 MeV to the triplet GS $^3P_0$ of $^{16}$O in the p$^{15}$N channel with one bound FS, for considering the total cross section of the radiative proton capture on $^{15}$N to the GS, as in the previous works [4-9]. There are two resonances at the energies up to 1.1 MeV, which can be compared with the $^3S_1$ scattering wave.

1. The first resonance is at the energy of 335(4) keV with the width of 110(4) keV

in the laboratory system (l.s.) and has the moment $J^\pi, T = 1^-0$ (see Table 16.22 in work [24]) – it can be caused by the triplet $^3S_1$ scattering state.

2. The second resonance is at the energy of 710(7) keV with the width of 40(40) keV (l.s.) and has the moment $J^\pi, T = 0^-1$. It can be caused by the $^1S_0$ scattering state, however there is no resonance in the capture cross section for it.

3. The third resonance is at the energy of 1028(10) keV with the width 140(10) keV (l.s.) and has the moment $J^\pi, T = 1^-1$ (see Table 16.22 in work [24]) – it also can be caused by the triplet $^3S_1$ scattering state.

These levels have the next energies in the center-of-mass system (c.m.): The first resonance is at the energy of 312(2) keV with the width of 91(6) keV – it corresponds to the excited state (ES) of $^{16}$O at the energy of 12.440(2) MeV. The third resonance is at the energy of 962(8) keV with the width of 130(5) keV – it corresponds to the ES of $^{16}$O at the energy of 13.090(8) MeV (see Table 16.13 in work [24]). We do not consider the second resonance, because, as it was mentioned, it does not give obvious contribution to the total radiative capture cross sections.

Carrying out the calculations of the total radiative capture cross sections the nuclear part of the p$^{15}$N intercluster potential is usually presented in the Gaussian form [4-9]

$$V(r) = -V_0 \exp(-\alpha r^2). \tag{6}$$

Immediately note that for the potential of the resonance $^3S_1$ waves without FSs two potentials were obtained, which correspond to two resonances of different width at different energies of 335 keV and 1028 keV. The parameters of the first potential:

$$V_{S1} = 1.0857 \text{ MeV}, \quad \alpha_{P1} = 0.003 \text{ fm}^{-2} \tag{7}$$

lead to the resonance energy of 335.0(1) keV and its width of 138(1) keV (l.s.), and the scattering phase shift is shown in Fig. 1 by the solid line. The relative accuracy of calculation of the $^3S_1$ scattering phase in these calculations is equal to $10^{-3}$ approximately and for the energy of 335 keV this potential leads to the value of the phase shift of 90.0(1)°. The next parameters were obtained for the second potential of the $^3S_1$ wave

$$V_{S1} = 105.059 \text{ MeV}, \quad \alpha_{P1} = 1.0 \text{ fm}^{-2}. \tag{8}$$

It leads to the resonance at 1028.0(5) keV and its width is equal to 140(1) keV (l.s.), and the scattering phase shift is shown in Fig. 1 by the dashed line – for resonance energy this potential also leads to the phase shift value of 90.0(1)°.

Here, we must draw attention [4-9] that the potential with the given number of BSs is constructed completely unambiguously using the energy values of the resonance level in spectra of $^{16}$O and its width. It is impossible to find other parameters $V_0$ and $\alpha$, which can be able correctly reproduce the level resonance energy and its width, if the specified number of FSs and ASs are given, which in this case equals zero. The depth of this potential unambiguously determines the location of the resonance, i.e., the level resonance energy $E_r$, and its width gives specific width $\Gamma_r$ of this resonance state. The error of parameters of such potential is determined by the measurement error of the level $E_r$ and its width $\Gamma_r$ [4-9]. However, here we must note that it is impossible, for the present, to construct the unified $^3S_1$ potential, which would contain two, noted above, resonances with different energies and widths. Therefore, the calculated total cross section for these resonances will consist of two parts – the first with the potential (7) and the second for the interaction (8), also both of these parts are the same $E$1 transition from the $^3S_1$ scattering wave to the $^3P_0$ GS of $^{16}$O.

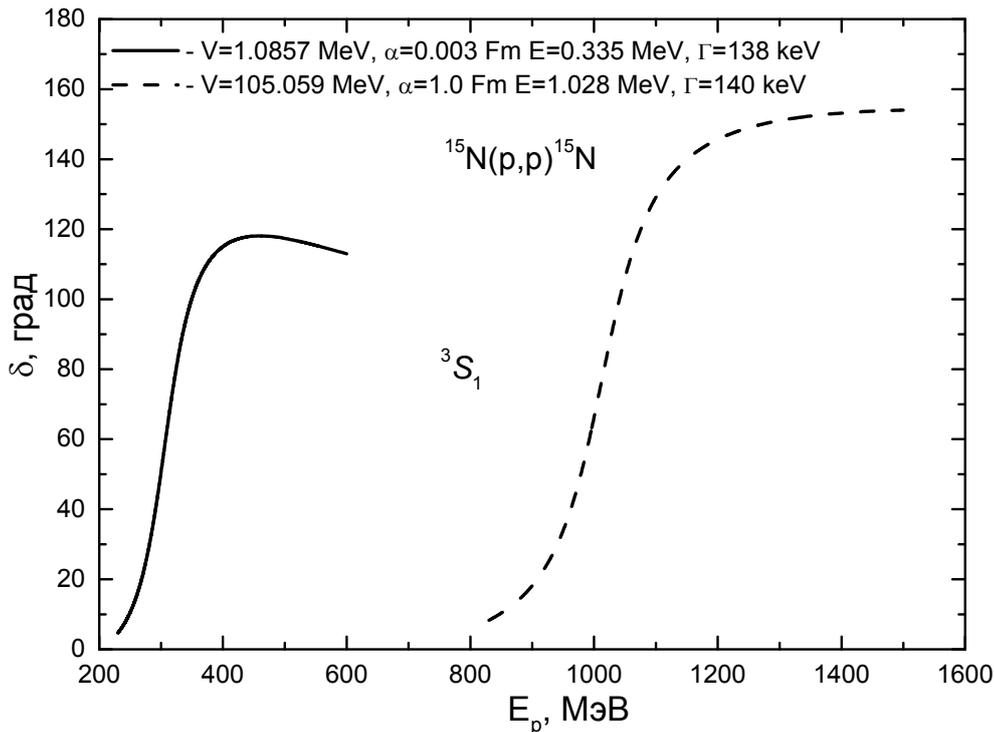

Fig. 1. The phase shifts of the p$^{15}$N elastic scattering in the $^3S_1$ wave

There are no resonance levels with $J = 0^+, 1^+, 2^+$ and widths more than 10 keV (see Table 16.22 in work [24]) in the spectra of the elastic p$^{15}$N scattering at the energies up to 1.05 MeV. Therefore, it is possible to use the parameter values for the potentials of nonresonance $^3P$ waves with one bound FS based on the assumption that in the considered energy range, i.e., up to 1.5 MeV, their phase shifts are equal to zero. The next parameters were obtained for such potentials

$$V_P = 14.4 \text{ MeV}, \quad \alpha_P = 0.025 \text{ fm}^{-2} \qquad (9)$$

The calculation of the $P$ phase shifts with this potential at the energy up to 1.5 MeV leads to their values from 180° to 179°. The unified Levinson theorem [10-12] was used for determination of the phase shift values at zero energy, therefore phase shifts of the potential with one bound FS have to start from 180°. There is the resonance with $J = 2^+$ [24] in the spectrum of $^{16}$O at the energy of 1.050(150) MeV, however the width values are not given for it, and the next $J = 1^+$ resonance with the width of 68(3) keV is located higher than 1.5 MeV.

Furthermore, we will construct the potential with FS of the $^3P_0$ state, which has to correctly reproduce the binding energy of the GS of $^{16}$O with $J^\pi T = 0^+0$ in the p$^{15}$N channel at -12.1276 MeV [24] and reasonably describe the mean square radius of $^{16}$O, which experimental value equals 2.710(15) fm [24], at the experimental radius of $^{15}$N – 2.612(9) fm [25]. The charged and the mass radius of proton are equal to 0.8775(51) fm [36] were used in these calculations. Consequently, the next parameters for the potential of the GS of $^{16}$O in the p$^{15}$N channel were obtained:

$$V_{G.S.} = 1057.99470 \text{ MeV}, \quad \alpha_{G.S.} = 1.2 \text{ fm}^{-2}. \qquad (10)$$

The potential leads to the binding energy of -12.12760 MeV at the FDM accuracy of $10^{-5}$ MeV, the mean square charged radius of 2.52 fm and the mass radius of 2.57 fm. The value of 1.94(1) at the range of 2–10 fm was obtained for the AC, written in the nondimensional form [40] $\chi_L(R) = \sqrt{2k_0}\, C_0\, W_{-\eta L+1/2}(2k_0R)$. The error of the constant is determined by its averaging over the noted above range interval. The phase shift of this potential is in the range from 180º to 179º at the energy up to 1.5 MeV.

The value of 192(26) fm$^{-1}$ for this AC is given in work [41] that after division on the spectrofactor value of 2.1 and the antisymmetrization coefficient of 16 [42] gives 5.71(77) fm$^{-1}$ or 2.39(88) fm$^{-1/2}$. Its recalculation to the nondimensional value at $\sqrt{2k_0} = 1.22$ gives 1.96(72) – this value is in a good agreement with the constant obtained above. The recalculation of the AC to the nondimensional value is needed because another definition of the AC were used in these works, that is

$$\chi_L(R) = CW_{-\eta L+1/2}(2k_0R), \qquad (11)$$

which differs from using here by factor $\sqrt{2k_0}$. In addition, the AC from [41] contains the spectrofactor equals, evidently, 2.1 and the antisymmetrization coefficient, obtained in the review [43].

Give here the second variant of the GS potential with the similar parameters, but slightly another AC

$$V_{G.S.} = 976.85193 \text{ MeV}, \quad \alpha_{G.S.} = 1.1 \text{ fm}^{-2}. \qquad (12)$$

It leads to the same binding energy, does not change the values of mean square charged and mass radii, and for the AC in the nondimensional form [40] at the range of 2–9 fm the value of 2.05(1) was obtained. The phase shift of this potential at the energy up to 1.5 MeV lies at the same interval of values as for the potential (10).

Earlier it was shown once and again [4-9] that the additional control of calculation of the binding energy GS or BS on the basis of the variational method (VM) [37,38] leads to the results coinciding with the FDM with the given determination accuracy of the binding energy of two-cluster system, therefore here we already not used the VM for the verification of this binding energy.

## 4. The total neutron capture cross sections on $^{15}$N

Going to the direct consideration of the results of the $M1$ and the $E1$ transitions to the GS of $^{16}$O, note that we succeeded [44,45,46] to find the experimental data for the total cross section of the process of the proton capture on the $^{15}$N in the energy range from 80 keV up to 1.5 MeV, which will be considered furthermore – these results are shown in Figs. 2a and 2b. The first part of the cross section of the $E1$ transition $^3S_1 \to {}^3P_0$ to the GS, calculated with the potentials of the scattering state (7) and the GS (10), is shown in Fig. 2a by the dashed line, and the dotted line shows the cross section of the $E1$ transition for the scattering potential (8) and the GS (10). In addition, we considered the $M1$ transition of the form $^3P_1 \to {}^3P_0$ with the potentials of the scattering state (9) and the GS (10). The results of these calculations are shown in Fig. 2a by the dot-dashed line and the total summed cross section for the referred above capture processes to the ground state is shown by the solid line. The $S$-factor is equal to 39.50(5) keV b and practically constant at the energy range of 50–60 keV. The similar calculations are given in Fig. 2b for the GS potential (12), these results differ from the previous one only in the low energy range. This potential of the GS leads to the $S$-factor values of 43.35(5) keV b at the energy range of 50–60 keV. The values of the $S$-factor at these energies, owing to their weak changes, can be considered, evidently, as the $S(0)$-factor for zero energy.

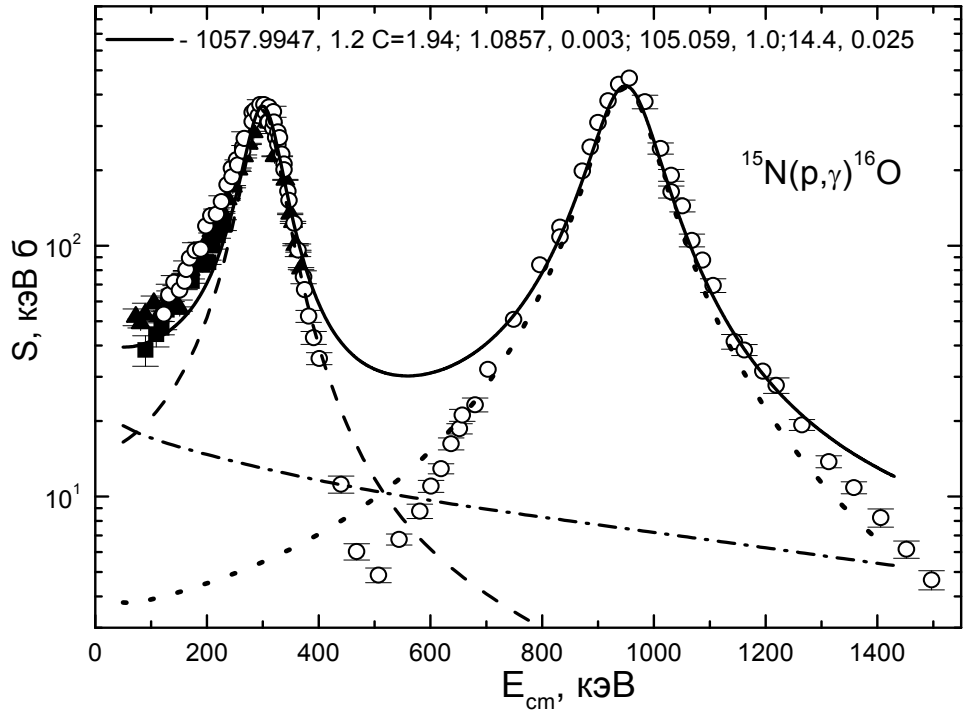

Fig. 2a. The astrophysical $S$-factor of the proton radiative capture by $^{15}$N on the GS. The experimental data: ▲ – [44], ■ – [45], ○ – [46]. Lines – the calculation of the total cross sections for transitions to the GS.

It can be noted that if the parameters of the resonance $^3S_1$ potential are fixed according to the resonance of phase shift relatively unambiguously and for the bound state they are chosen on the basis of the description of the bound state characteristics, therefore for the $^3P_1$ potential with the FS (9) which leads to zero scattering phases, another parameter values are possible. However, only the given above parameters (9) allow one to obtain the acceptable results of calculations for the transition from the $^3P_1$ wave to the GS, which are shown in Fig. 2a and 2b by the dot-dashed lines. Now, it is impossible to draw a unique conclusion about the form and depth of such scattering potential.

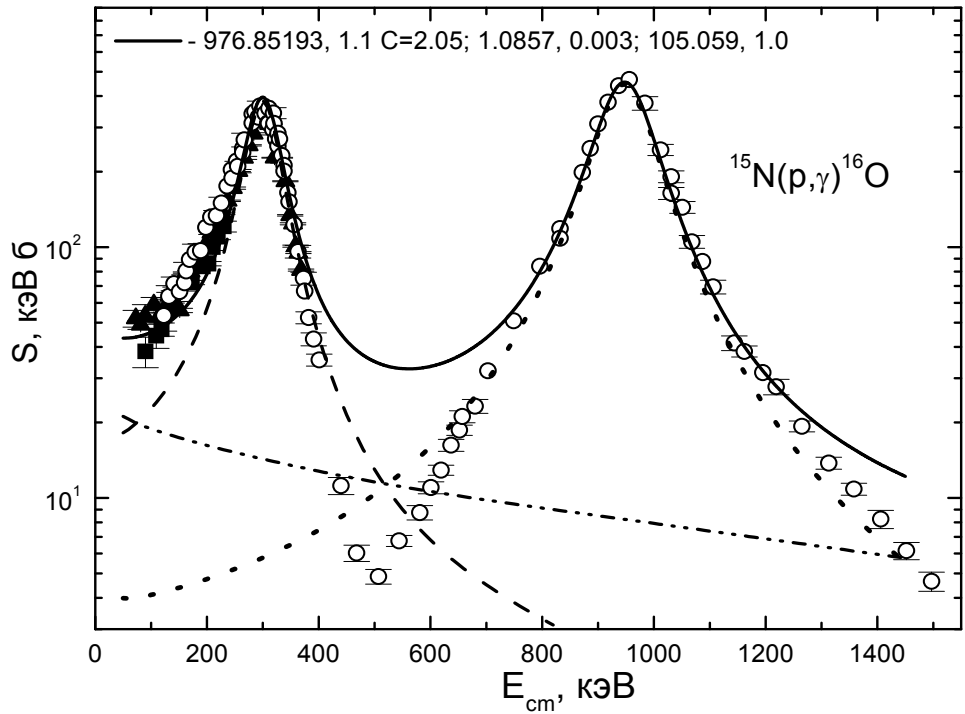

Fig. 2b. The same as in Fig. 2a.

In addition, the medium-energy region, where the minimum of cross sections at 500 keV is observed, is described badly. It is connected with the inefficient abrupt decrease of the cross section after the first resonance at the energy above 400 keV (the dashed line in Fig. 2a and 2b) and with the inefficient abrupt rising of it before the second resonance at the energies up to 600–650 keV (the dotted line in Fig. 2a and 2b). It is also connected with the value of the cross section for the $M1$ transition, which is, for this energy region, more than minimal experimental value of 4.9 keV b at 507 keV [46]. And here, evidently, the results of the phase shift analysis of angular distribution of the elastic p$^{15}$N scattering are needed for the construction of the correct potentials of two resonances in the $^3S_1$ wave.

## 5. Conclusion

Thereby, the intercluster potentials of the bound state, constructed on the basis of the quite obvious requirements for description of the binding energy, the mean square radii of $^{16}$O and the AC values in the p$^{15}$N channel, and also the scattering potentials describing the resonances allow one to reproduce generally correctly the available experimental data for the total cross sections of the proton radiative capture on $^{15}$N at low energies [46]. Meanwhile, all p$^{15}$N potentials using here are constructed on the basis of the given above classification of the FSs and the ASs according to Young tableaux.

However, it is difficult to do defined and final conclusions if we have not results of the phase shift analysis, carried out on the basis of differential cross sections of the p$^{15}$N elastic scattering. Such data are needed, approximately, in the range of 0.3–1.5 MeV, however the results at the energies lower 1 MeV are absent up to now. Meanwhile, the available data on the differential cross sections [47] were obtained in the form of excitation function only at two angles and only above the energy of 0.97 MeV. Therefore, furthermore it is desirable to carry out the detailed measurement of the differential cross sections of the elastic scattering in the energy range from 0.1–0.3 MeV up to 1.3–1.5 MeV. Such data have to contain angular distributions in the range of two resonances at 335 keV and 1028 keV [24] at angles from 30º to 170º.

**Acknowledgments**

In conclusion, the authors express their deep gratitude to Akram Mukhamedzhanov for extremely useful discussion of the certain questions connected with the ACs, which were considered in this work.

The work was performed under the grant No. 0601/GF "The experimental and theoretical study of properties of the excited halo-states of neutron-rich nuclei $^9$Be, $^{11}$B, $^{13,15}$C and $^{15}$N" of the Ministry of Education and Science of the Republic of Kazakhstan.